\documentstyle[epsf]{ioplppt}

\newcommand{\vv}[1]{\mbox{\boldmath $#1$}}                 
\newcommand{\vvsmall}[1]{\mbox{\scriptsize\boldmath $#1$}} 
\newcommand{\gapr}{\raisebox{-.6ex}{\mbox{                 
$\stackrel{>}{\mbox{\scriptsize $\sim$}}\:$}}}             
\newcommand{\lapr}{\raisebox{-.6ex}{\mbox{                 
$\stackrel{<}{\mbox{\scriptsize $\sim$}}\:$}}}             
\newcommand{\ab}{a_{\rm B}}                   
\newcommand{\kb}{k_{\rm B}}                   
\newcommand{\mel}{m_{\rm e}}                      
\newcommand{\mpr}{m_{\rm p}}                      
\newcommand{\mH}{m_{\rm H}}                       
\newcommand{\be}{\begin{equation}}
\newcommand{\ee}{\end{equation}}
\begin{document}
\jl{2}
\title{Hydrogen atom moving across a strong magnetic field:
analytical approximations
}[
Hydrogen atom moving in strong magnetic field]

\author{A Y Potekhin\ftnote{1}{E-mail: palex@astro.ioffe.rssi.ru}}
\address{Ioffe Physico-Technical Institute,
Politekhnicheskaya 26, St Petersburg 194021, Russia
}

\begin{abstract}
Analytical approximations are constructed
for binding energies, quantum-mechanical sizes
and oscillator strengths of main radiative transitions
of hydrogen atoms arbitrarily moving in magnetic fields
$\sim10^{12}-10^{13}$~G.
Examples of using the obtained approximations
for determination of maximum transverse velocity
of an atom and for evaluation of 
absorption spectra in magnetic neutron star atmospheres 
are presented.
\end{abstract}

\pacs{32.10.Hq, 32.60.+i, 32.70.Cs, 97.60.Jd}

\vspace*{3cm}
\mbox{\protect\jpb {\bf 31} (1998) 49--63}

\maketitle
\setcounter{page}{49}
\section{Introduction}
An atom moving across magnetic field is equivalent to an atom placed
in perpendicular magnetic and electric fields. 
We consider the hydrogen atom moving in a magnetic
field $\vv{B}$, strong enough to significantly squeeze the electron 
wavefunction. Quantitatively, the parameter 
$\gamma=\hbar\omega_c /(2{\rm\,Ryd})=B/(2.35\times10^9{\rm\,G})$ 
is assumed large.
Here, $\omega_c  = eB/\mel c$ is the electron cyclotron
frequency, and ${\rm Ryd}=\mel e^4/2\hbar^2$ is the ground-state energy
of the field-free atom.

Although only small values of $\gamma$ are
available in laboratory, large values are not uncommon in
astrophysics.
Spectra of some white-dwarf stars have been interpreted
as produced by hydrogen at field strengths between 
$10^6$~G and $10^9$~G (Wunner and Ruder 1987, Fassbinder and 
Schweizer 1996, and references therein). 
Neutron stars which are observed as radio pulsars
reveal field strengths
in excess of $2\times10^8$~G, and  over
half of them possess magnetic fields from $10^{12}$~G 
to $2\times10^{13}$~G (Taylor \etal 1993).
Absorption of radiation by strongly magnetized atomic 
hydrogen may have large effects
on ultraviolet and X-ray spectra of the neutron stars,
which are measured with modern space telescopes 
(Pavlov \etal 1995).

The physics of solid state presents another important field of
application of quantum-mechanical calculations of strongly
magnetized hydrogen atoms. Excitons and shallow impurities in
semiconductors reveal hydrogen-like spectra with scaled values 
of $\omega_c $ and Ryd. Such scaling offers a possibility to
reach the regime $\gamma>1$ in an experiment (e.g.,
Elliott and Loudon 1960,
Klaassen \etal 1997).

The non-moving hydrogen 
atom in a strong magnetic field was thoroughly studied
in the past two decades (Ruder \etal 1994, and references therein).
Extensive tables of binding 
energies have been presented by R\"osner \etal (1984)
and supplemented by 
Wintgen and Friedrich (1986), Ivanov (1988),
Xi \etal (1992) and Kravchenko \etal (1996).
Tables of oscillator strengths 
at various values of $\gamma$ have been published 
by Forster \etal (1984); analytical fits to
photoionization cross sections
at $\gamma\gg1$ have been proposed by Potekhin and Pavlov (1993).

The problem of a moving atom (or an atom in crossed fields)
is much more complicated because of the absence of
the axial symmetry. Much work has been done at low field strengths
(e.g., Melezhik 1993), where the second-order perturbation theory was
applicable (e.g., Braun and Solov'ev 1984, and references therein).
The simplifying approximation of infinite proton mass, exploited in
this regime, breaks down in strong fields because of collective
motion effects studied in detail by Avron \etal (1978),
Baye and Vincke (1990) and Dippel \etal (1994).
In particular, so-called decentred states 
(with the electron localized mostly in the ``magnetic well'' aside 
from the Coulomb centre) are likely to be populated.
These exotic states have been predicted by Burkova \etal (1976)
and studied by Ipatova \etal (1984), Baye \etal (1992),
Dzyaloshinskii (1992) and Schmelcher (1993).
As well as the usual ``centred'' states, the decentred states have
an infinite discrete energy spectrum 
(Potekhin 1994, hereafter Paper~I).
Collective-motion effects on the centred states
of the strongly magnetized hydrogen atom
have been considered by Vincke and Baye (1988) 
and Pavlov and M\'esz\'aros (1993)
in frames of the theory of perturbation.

Completely non-perturbative results, covering both centred and 
decentred
states as well as the transition region, were first presented by
Vincke \etal (1992) for binding energies and wavefunctions. 
In Paper~I, additionally, oscillator strengths have been considered.
Pavlov and Potekhin (1995, hereafter Paper~II) 
studied spectral line shapes, and
Potekhin and Pavlov (1997) calculated 
photoionization cross sections.
None of these numerical results, however, has been published 
in an easy-to-use form of tables or analytical expressions. 
This paper provides such expressions for the magnetic field
strengths typical of neutron stars, $300\leq\gamma\leq10^4$.
This range is physically distinguished, since at weaker fields
the transition region is strongly complicated by multiple
narrow anticrossings (Vincke \etal 1992).
The relative simplicity of the spectrum at 
$\gamma\gapr300$
facilitates analytical description. The upper bound, $\gamma\sim10^4$,
corresponds to the onset of non-negligible relativistic effects
(Chen and Goldman 1992).

In the next section we recall the basic definitions and physical 
properties of
a hydrogen atom arbitrarily moving in a strong magnetic field.
In Section 3, we first 
present accurate analytical fits to binding energies, 
depending
on the state of motion, for a number of 
bound states and various field strengths. 
Then we derive analytical approximations 
continuously depending on $\gamma$. 
As a by-product, simple and accurate approximations are obtained 
for binding energies of the non-moving atom at any $\gamma\gapr1$.
The obtained formulae are applied 
to evaluation of the maximum transverse velocity 
of the strongly magnetized atom.
Section \ref{sect-geom} is devoted to analytical approximations of 
quantum-mechanical sizes
and main oscillator strengths of the atom. 
In section \ref{sect-shape}, an example of using the obtained
expressions for calculation of absorption coefficients of strongly 
magnetized, hot hydrogen plasma is presented.

\section{Centred and decentred states: 
general description}
Motion of the hydrogen atom in a magnetic field 
can be conveniently described by the pseudomomentum 
(e.g., 
Johnson \etal 1983)
\be 
\vv{K} = \mpr \dot{\vv{r}}_{\rm p} + \mel \dot{\vv{r}}_{\rm e}
 - \frac{e}{c} \vv{B} \times 
(\vv{r}_{\rm e} -\vv{r}_{\rm p}),
\ee 
where the subscript $i=$e or $i=$p indicates electron 
or proton, respectively, 
\be 
   \dot{\vv{r}}_i={{\rm i}\over\hbar}[H_{\rm tot},\vv{r}_i]
   = -{{\rm i}\hbar\over m_i}\nabla_i 
   -{q_i\over m_i c}\vv{A}(\vv{r}_i)
\ee 
is the velocity operator, $m_i$ the mass, 
$q_{\rm p}=-q_{\rm e}=e$ the charge, 
$\vv{A}(\vv{r})$ the vector potential of the field,
and $H_{\rm tot}$ the two-particle Hamiltonian operator. 
Gorkov and Dzyaloshinskii (1968) have shown that
in the representation in which all three components of $\vv{K}$ 
have definite values the pseudoseparation of 
the centre-of-mass motion can be performed, 
that is, the relative motion 
can be described in terms of a one-particle Hamiltonian
which depends on $\vv{K}$. 
The expectation value
of the velocity of the atom is $\nabla_{\vvsmall{K}} E$,
where $E$ is the expectation value of the energy.

It is convenient to describe the centred states 
of the atom 
using the relative coordinate
$\vv{r}^{(0)}=\vv{r}_{\rm e}-\vv{r}_{\rm p}$ 
as independent variable and the 
axial gauge of the vector potential, 
$\vv{A}(\vv{r}) = \frac{1}{2}\vv{B}\times \vv{r}$.
For the decentred states, the ``shifted''
representation 
(Gorkov and Dzyaloshinskii 1968) is more convenient.
In the latter representation, 
the independent variable is 
$\vv{r}^{(1)}=\vv{r}_{\rm e}-\vv{r}_{\rm p}-\vv{r}_c$
and the gauge is
$\vv{A}(\vv{r}) = \frac{1}{2}\vv{B}\times 
(\vv{r}-\left[(\mpr-\mel)/\mH\right]\vv{r}_c)$. 
Here, 
$\vv{r}_c= (c/eB^2)\,\vv{B}\times\vv{K}$
is the relative guiding centre and
$\mH=\mpr+\mel$ is the mass of the atom.

Let us assume that $\vv{B}$ is directed 
along the $z$-axis. The $z$-component of the pseudomomentum 
corresponding to the motion
along the field separates exactly from the Hamiltonian, giving the
kinetic term $K_z^2/2\mH$,
while the transverse components $\vv{K}_\perp$
produce non-trivial effects.
Therefore we assume $K_z=0$ 
and $\vv{K}_\perp=\vv{K}$ hereafter.

If there were no Coulomb attraction, then
the electron Landau number $n=0,1,2,\ldots$ 
and the $z$-projection $s$ of
the angular momentum of the relative motion 
would be exact quantum numbers
(since $\vv{K}$ is definite, the electron and proton 
do not possess definite $z$-projections of the angular momenta 
separately from each other --- see Johnson \etal\ 1983). 
In this case the transverse part of the wavefunction 
could be described by a Landau function
$\Phi_{ns}(\vv{r}^{(1)}_\perp)$, where
$\vv{r}^{(1)}_\perp$ is the projection of $\vv{r}^{(1)}$
in the $(xy)$-plane
and $s$ is defined in the shifted reference frame (e.g., Paper~I).
The energy of the transverse excitation 
(with the zero-point and spin terms subtracted) is
\be 
   E^\perp_{ns} = [
   n+(\mel/\mpr)(n+s)]
\hbar\omega_c.
\ee 

A wavefunction $\psi_\kappa$ of an atomic state 
$|\kappa\rangle$ can be expanded over the complete set of the 
Landau functions
\be 
   \psi_\kappa^{(\eta)}(\vv{r}^{(\eta)}) = 
   \sum_{ns} \Phi_{ns}(\vv{r}_\perp^{(\eta)})\, 
   g^{(\eta)}_{n,s;\kappa}(z),
\label{expan}
\ee 
where $\eta=0$ or 1 indicates the 
conventional or shifted representation,
respectively (a generalization to continuous $\eta$ 
in Paper~I proved to be less useful).
The adiabatic approximation used in early works
(Gorkov and Dzyaloshinskii 1968, Burkova \etal 1976)
corresponds to retaining
only one term in this expansion.

A bound state can be numbered as 
$|\kappa\rangle = |ns\nu\vv{K}\rangle$, 
where $n$ and $s$ relate to the leading term of the 
expansion (\ref{expan}), and $\nu$ enumerates longitudinal 
energy levels
\be 
   E^\|_{ns\nu}(K) = 
   E_\kappa - E^\perp_{ns}
\label{elong}
\ee 
and controls the $z$-parity:
$g^{(\eta)}_{n,s;\kappa}(-z)=(-1)^\nu g^{(\eta)}_{n,s;\kappa}(z)$. 
This way of numbering is conventional for the non-moving atom
at $\gamma\gapr1$. The states $\nu=0$ are tightly bound 
in the Coulomb well, while the states $\nu\geq1$ 
are hydrogen-like, with binding energies below 1 Ryd.
For a moving atom, this way of 
numbering remains unambiguous at $\gamma\gapr 300$,
in spite of the fact that there may not exist an
obvious leading term of (\ref{expan})
in this case  (Paper~I).

The inequality $E_\kappa<0$
determines truly bound states, as opposed to 
the ones subject to autoionization.
In particular, all states with $n\neq0$ 
belong to continuum at $\gamma\gapr0.2$ and will not
be considered hereafter.

Since the transverse factors $\Phi_{ns}$ in (\ref{expan})
are known analytically, only the one-dimensional longitudinal
functions $g^{(\eta)}_{ns;\kappa}$ are to be found
numerically. An algorithm which is most efficient at $\gamma\gg1$
has been described in Paper~I. 
At small pseudomomenta $K$,
the states $\nu=0$ remain tightly bound and centred,
the average electron-proton displacement
$\bar{x}$ being considerably smaller than $r_c$.
For the hydrogen-like states $\nu\geq 1$, however,
$\bar{x}$ is close to $r_c$ at any $K$.

According to the second-order perturbation approximation
at small $K$, the absolute expectation value of the velocity 
$v=\partial E_\kappa/\partial K$ 
in a bound state $|\kappa\rangle$ equals  
$K/M^\perp_{ns\nu}$, 
where $M^\perp_{ns\nu}$ is the effective 
``transverse'' mass (Vincke and Baye 1988, 
Pavlov and M\'{e}sz\'{a}ros 1993).
$M^\perp_{ns\nu}$ always exceeds $\mH$, and it is the greater
the stronger the field and the higher the
considered atomic level.

The larger $K$, the greater is the distortion of 
the wavefunction towards $\vv{r}_c$,
caused by the motion-induced electric field
in the co-moving reference frame. 
The perturbation approximation becomes increasingly
inaccurate, until
near some critical value $K_c$
a transition to the decentred state occurs,
and the character of the motion totally changes.
With further increasing $K$,
the transverse velocity
decreases and tends to zero, 
while the electron-proton separation 
increases and tends to $r_c$. 
Thus, for the decentred states, the pseudomomentum 
characterizes electron-proton separation
rather than velocity.

The shifted ($\eta=1$) adiabatic approximation becomes fairly
good at $K\gg K_c$.
At very large $K$ the longitudinal functions
become oscillator-like, corresponding to
a wide, shallow parabolic potential well
of a depth
 $\simeq e^2/r_c$ (Burkova \etal 1976).
For a fixed $\nu$, this limit is reached at 
$K\gg(\nu+\case12)^2 \hbar/\ab$, where $\ab$ is the Bohr radius.
Still at arbitrarily large $K$ there remain infinite number
of bound states with high values of $\nu$
whose longitudinal wavefunctions are goverened
by the Coulomb tail rather than by the parabolic core
of the effective one-dimentional potential (Paper~I). 

The decentred states of the atom 
at $K>K_c\sim10^2$~au have relatively low binding
energies and large quantum-mechanical sizes,
$l\sim K/\gamma$~au; therefore
they are expected to be destroyed 
by collisions with surrounding particles
in the laboratory and in the white-dwarf atmospheres.
In neutron-star atmospheres
at $\gamma\gapr10^3$, however, the decentred states
may be significantly populated (Paper~II).
This necessitates inclusion of the entire
range of $K$ below and above $K_c$
in the consideration.

\section{Binding energies}
\subsection{Dependence of the energies on the pseudomomentum
at selected field strengths}
\label{sect-1dim}
We have calculated binding energies
of the hydrogen atom moving across the strong magnetic field
at $\gamma=300$, 600, 1000, 2000, 3000 and 10\,000 
for several lowest tightly-bound
and hydrogen-like states, using the technique described in Paper~I.
At each value of $\gamma$ and for each state,
the calculations have been performed at $K=0$ and at
about 50--100 values of $K$ from $K\leq10$~au
to $K\geq10^4$~au, approximately
equidistant in $\log K$ but with additional 
points near avoided crossings.
The calculated energies 
have accuracy of 3--5 digits.

In applications, however, one usually has to deal
with a distribution of atoms over 
more or less broad band of values of 
the pseudomomentum $K$, and to calculate the observable quantities
by averaging over $K$.
This makes it highly desirable to have
an analytical approximation of the $K$-dependence
of the energies, $E(K)$. 
Lai and Salpeter (1995) were the first to present 
an analytical fit to
$E(K)$, which was rather accurate for the ground state 
at $K<K_c$
but could not be applied to excited or decentred states.

We describe the longitudinal 
energy (\ref{elong}) by the formula
\be 
   |E^\|_{ns\nu}(K)| = 
   {E_{ns\nu}^{(1)}(K) \over 1+(K/K_c)^{1/\alpha}} 
   + {E_{ns\nu}^{(2)}(K) \over 1+(K_c/K)^{1/\alpha}}.
\label{eappr0}
\ee 
The two-term structure of (\ref{eappr0}) is dictated by
the necessity to describe the two physically distinct
regions of $K$ below and above $K_c$.
The parameter $\alpha$ has the meaning of the width 
of the transition region near $K_c$
in logarithmic scale of pseudomomenta. 
As noted in Paper~I, for the tightly-bound states
$K_c$ is close to $(2\mH E^{(0)}_{ns\nu})^{1/2}$, 
where $E^{(0)}_{ns\nu}\equiv -E_{ns\nu}^\|(0)$. 
We write $K_c=q_0(2\mH E^{(0)}_{ns\nu})^{1/2}$
and treat $q_0$ as a fitting parameter.

Intricate structure
of the region of avoided crossings (see Vincke \etal 1992)
complicates its accurate analytical description. 
We have chosen to keep our formulae simple
at cost of decreasing accuracy near these crossings. 

For the tightly-bound states, we parametrize 
the functions $E^{(j)}(K)$ as follows:
\begin{eqnarray}
   E_{0s0}^{(1)}(K) &=& E^{(0)}_{0s0}-
   {K^2\over 2m_{\rm eff}+q_1 K^2/E^{(0)}_{0s0} },
\label{eappr1}
\\
   E_{0s0}^{(2)}(K) &=& 
   2 \left[r_\ast^2+r_\ast^{3/2}+q_2 r_\ast \right]^{-1/2}{\rm~Ryd},
\label{eappr2}
\end{eqnarray}
where $r_\ast =r_c/\ab=K/(\gamma$~au),
$q_1$ and $q_2$ are dimensionless fitting parameters,
and $m_{\rm eff}$ is the effective mass
which is close to (but not necessarily coincident with)
the transverse effective mass $M^\perp_{ns\nu}$
obtained by the perturbation technique.

In the considered range of $\gamma$,
the parameter $q_1$ can be approximated as 
\[
   q_1=\left\{
   \begin{array}{ll}
   \lg(\gamma/300) & \mbox{if $s=0$,} \\
   0.5 & \mbox{otherwise.}
   \end{array}\right.
\]
Optimal values of the other parameters are listed in \tref{tab1}.
The last column presents the root-mean-square (rms)
difference $\sigma_E$ 
between the computed and fitted energies.
Maxmum errors of the fit ($\lapr3 \sigma_E$) 
occur near the avoided crossings.

\begin{table}
\caption{Parameters of the analytical approximation 
\protect\eref{eappr0}--\protect\eref{eappr2} for the energies
of tightly-bound states $|0s0\rangle$}
\lineup
\label{tab1}
\begin{indented}
\item[]\begin{tabular}{@{}lrcclllc}
\br
$s$ & $\gamma$\0\0 & $E^{(0)}$ (Ryd) & $\lg(m_{\rm eff}/\mH)$ & 
$q_0$ & $\alpha$ & $q_2$ & $\sigma_E$ (Ryd) \\
\mr
0 &     300 &   10.722 &  0.009  &  0.859 &  0.001 &  0.102 &  0.028\\
  &     600 &   13.210 &  0.042  &  0.811 &  0.107 &  0.157 &  0.040\\
  &    1000 &   15.325 &  0.072  &  0.823 &  0.117 &  0.189 &  0.025\\
  &    2000 &   18.610 &  0.141  &  0.850 &  0.178 &  0.233 &  0.018\\
  &    3000 &   20.770 &  0.175  &  0.873 &  0.191 &  0.244 &  0.017\\
  &   10000 &   28.286 &  0.319  &  1.019 &  0.173 &  0.275 &  0.027\\
\mr
1 &     300 & \0 7.669 &  0.161  &  0.963 &  0.132 &  0.115 &  0.026\\
  &     600 & \0 9.607 &  0.269  &  1.060 &  0.093 &  0.160 &  0.021\\
  &    1000 &   11.277 &  0.369  &  1.147 &  0.060 &  0.176 &  0.024\\
  &    2000 &   13.904 &  0.578  &  1.195 &  0.122 &  0.215 &  0.016\\
  &    3000 &   15.649 &  0.701  &  1.202 &  0.147 &  0.235 &  0.014\\
  &   10000 &   21.830 &  0.944  &  1.337 &  0.298 &  0.240 &  0.033\\
\mr
2 &     300 & \0 6.450 &  0.304  &  1.184 &  0.030 &  0.120 &  0.017\\
  &     600 & \0 8.142 &  0.497  &  1.197 &  0.081 &  0.181 &  0.014\\
  &    1000 & \0 9.610 &  0.643  &  1.262 &  0.074 &  0.195 &  0.014\\
  &    2000 &   11.937 &  0.931  &  1.291 &  0.127 &  0.230 &  0.014\\
  &    3000 &   13.493 &  1.093  &  1.320 &  0.153 &  0.240 &  0.022\\
\mr
3 &     300 & \0 5.734 &  0.466  &  1.263 &  0.039 &  0.122 &  0.015\\
  &     600 & \0 7.274 &  0.701  &  1.273 &  0.082 &  0.183 &  0.012\\
  &    1000 & \0 8.617 &  0.897  &  1.347 &  0.090 &  0.204 &  0.018\\
  &    2000 &   10.755 &  1.252  &  1.403 &  0.131 &  0.232 &  0.019\\
  &    3000 &   12.191 &  1.451  &  1.457 &  0.154 &  0.240 &  0.026\\
\mr
4 &     300 & \0 5.243 &  0.616  &  1.330 &  0.050 &  0.128 &  0.013\\
  &     600 & \0 6.676 &  0.892  &  1.342 &  0.095 &  0.194 &  0.011\\
  &    1000 & \0 7.929 &  1.124  &  1.437 &  0.096 &  0.211 &  0.017\\
  &    2000 & \0 9.933 &  1.555  &  1.544 &  0.114 &  0.229 &  0.016\\
\mr
5 &     300 & \0 4.877 &  0.755  &  1.391 &  0.058 &  0.128 &  0.012\\
  &     600 & \0 6.227 &  1.086  &  1.393 &  0.107 &  0.199 &  0.012\\
  &    1000 & \0 7.413 &  1.354  &  1.545 &  0.130 &  0.229 &  0.010\\
\mr
6 &     300 & \0 4.589 &  0.888  &  1.448 &  0.062 &  0.123 &  0.013\\
  &     600 & \0 5.874 &  1.281  &  1.441 &  0.121 &  0.207 &  0.013\\
  &    1000 & \0 7.004 &  1.668  &  1.587 &  0.107 &  0.210 &  0.018\\
\mr
7 &     300 & \0 4.355 &  1.021  &  1.504 &  0.070 &  0.132 &  0.013\\
  &     600 & \0 5.585 &  1.480  &  1.473 &  0.139 &  0.213 &  0.014\\
\br
\end{tabular}
\end{indented}
\end{table}

Binding energies of the hydrogen-like states
are approximated by the same formula \eref{eappr0}
but with slightly different expressions
for $E^{(1)}$ and $E^{(2)}$. For these states,
$M^\perp_{ns\nu}$ exceeds $\mH$ by orders of magnitude,
and the perturbation method fails already 
at small values of $K$ 
(Pavlov and M\'{e}sz\'{a}ros 1993), 
which renders the notion of transverse mass 
practically useless for the fitting. 
Thus we consider $m_{\rm eff}$ as effectively infinite
and put $E^{(1)}_{0s\nu}=E^{(0)}_{0s\nu}$ ($\nu\geq1$).
Furthermore, the transition region is not well defined,
and therefore $K_c$ and $\alpha$ lose
their clear meaning and become mere
fitting parameters. 

The function $E^{(2)}(K)$ that describes the longitudinal energy
at large $K$ is now
\be 
      E_{0s\nu}^{(2)}(K) =
   \left\{(2{\rm~Ryd})^{-1}\left[
   r_\ast^2+(2\nu+1)r_\ast^{3/2}+q_2 r_\ast 
   \right]^{1/2}
   +1/E^{(0)}_{0s\nu}\right\}^{-1}\!,
\label{eappr3}
\ee 
where $r_\ast $ and $E^{(0)}$ have the same meaning as before.
The first and second terms in the square brackets ensure
the correct asymptotic behaviour (Paper~I).
In this case,
\[
   q_2=\left\{
   \begin{array}{ll}
   \nu^2-1 & \mbox{(odd $\nu$)}\\
   \nu^2+2^{\nu/2}\lg(\gamma/300) \qquad & \mbox{(even $\nu$).}
   \end{array}\right.
\]
Optimal values of the parameters $q_0$ and $\alpha$
are listed in \tref{tab2}.
As well as in \tref{tab1}, the last column presents rms errors
which are several times smaller than the maximum errors
near anticrossings.

\begin{table}
\caption{Parameters of the analytical approximation 
\protect\eref{eappr0}, \protect\eref{eappr3} for the energies
of hydrogen-like states $|0s\nu\rangle,~\nu\geq1$.}
\label{tab2}
\lineup
\begin{indented}
\item[]\begin{tabular}{@{}llrclll}
\br
$s$ & $\nu$ & $\gamma$\0\0 & $E^{(0)}_{0s\nu}$ (Ryd) & $q_0$ &
$\alpha$ & $\sigma_E$ (Ryd) \\
\mr
0 & 1 &  300 &   0.9643 & 1.751 & 0.7081 & 0.0013 \\
  &   &  600 &   0.9781 & 3.019 & 0.7441 & 0.0013 \\
  &   & 1000 &   0.9850 & 4.595 & 0.7604 & 0.0018 \\
  &   & 2000 &   0.9912 & 8.467 & 0.7977 & 0.0017 \\
  &   & 3000 &   0.9936 & 12.43 & 0.8095 & 0.0012 \\
  &   &10000 &   0.9976 & 39.65 & 0.8052 & 0.0023 \\
\mr
0 & 2 &  300 &   0.5522 & 1.064 & 0.6186 & 0.0006 \\ 
  &   &  600 &   0.5755 & 1.463 & 0.6252 & 0.0005 \\ 
  &   & 1000 &   0.5917 & 1.885 & 0.6322 & 0.0018 \\
  &   & 2000 &   0.6125 & 2.632 & 0.6255 & 0.0007 \\ 
  &   & 3000 &   0.6240 & 3.143 & 0.6406 & 0.0037 \\
  &   &10000 &   0.6554 & 4.810 & 0.6573 & 0.0022 \\
\mr
0 & 3 &  300 &   0.2456 & 5.608 & 0.8501 & 0.0005 \\ 
  &   &  600 &   0.2473 & 10.68 & 0.8495 & 0.0013 \\
  &   & 1000 &   0.2482 & 16.67 & 0.8617 & 0.0003 \\ 
  &   & 2000 &   0.2489 & 31.35 & 0.8940 & 0.0002 \\ 
  &   & 3000 &   0.2492 & 45.96 & 0.8966 & 0.0002 \\ 
  &   &10000 &   0.2498 & 150.1 & 0.8956 & 0.0003 \\ 
\mr
0 & 4 &  300 &   0.1814 & 2.145 & 0.7140 & 0.0025 \\
  &   &  600 &   0.1858 & 2.868 & 0.6699 & 0.0003 \\ 
  &   & 1000 &   0.1887 & 3.566 & 0.6609 & 0.0002 \\ 
  &   & 2000 &   0.1924 & 4.963 & 0.6165 & 0.0002 \\ 
  &   & 3000 &   0.1945 & 5.908 & 0.5970 & 0.0003 \\ 
  &   &10000 &   0.1999 & 8.965 & 0.5675 & 0.0006 \\ 
\mr
0 & 5 &  300 &  0.10982 & 10.05 & 0.9245 & 0.00014 \\
  &   &  600 &  0.11032 & 18.58 & 0.9422 & 0.00014 \\
  &   & 1000 &  0.11057 & 29.87 & 0.9404 & 0.00010 \\
  &   & 2000 &  0.11079 & 56.85 & 0.9630 & 0.00009 \\ 
  &   & 3000 &  0.11088 & 83.66 & 0.9619 & 0.00009 \\ 
  &   &10000 &  0.11104 & 273.3 & 0.9745 & 0.00006 \\ 
\mr
0 & 6 &  300 &  0.08920 & 2.435 & 0.8688 & 0.00054 \\
  &   &  600 &  0.09068 & 4.328 & 0.7156 & 0.00016 \\
  &   & 1000 &  0.09167 & 5.237 & 0.7205 & 0.00018 \\
  &   & 2000 &  0.09294 & 7.419 & 0.6593 & 0.00010 \\ 
  &   & 3000 &  0.09362 & 8.825 & 0.6237 & 0.00016 \\
  &   &10000 &  0.09542 & 13.43 & 0.5906 & 0.00026 \\
\mr
1 & 1 &  300 &   0.9407 & 2.109 & 0.6794 & 0.0010 \\ 
  &   &  600 &   0.9640 & 3.553 & 0.7029 & 0.0024 \\
1 & 2 &  300 &   0.5138 & 1.930 & 0.6417 & 0.0038 \\
2 & 1 &  300 &   0.9223 & 2.421 & 0.6553 & 0.0014 \\
\br
\end{tabular}
\end{indented}
\end{table}

In both tables \ref{tab1} and \ref{tab2}, only truly bound
(not autoionizing) states are considered. 
For example, all states with $s > 0$, $\nu>0$
belong to continuum at $\gamma > 673$, therefore
\tref{tab2} does not contain entries for them
at $\gamma=1000$ and higher. 
\subsection{Two-dimensional approximations}
\label{sect-2dim}
Equations (\ref{eappr0})--(\ref{eappr3}) 
help us to derive approximations of the 
binding energies as functions of two continuous arguments
$\gamma$ and $K$. For this purpose, we replace the 
numerical parameters listed in 
tables \ref{tab1} and \ref{tab2} by analytical functions of $\gamma$.

One of these parameters ---
the longitudinal energy of the atom at rest $E^{(0)}$ ---
has an independent significance.
For this reason, we have constructed
an accurate fit to $E^{(0)}$ 
in a possibly widest range of $\gamma$ values.
For the tightly-bound states,
we have
\be 
\fl
   E^{(0)}_{0s0}(\gamma)/\mbox{Ryd}=\ln\left(
   \exp\left[(1+s)^{-2}\right]
   +p_1\left[\ln(1+p_2\sqrt{\gamma})\right]^2\right)
   +p_3\left[\ln(1+p_4\gamma^{p_5})\right]^2.
\label{e0gamma}
\ee 
The parameters $p_1-p_5$ depend on $s$; they are presented in
\tref{tab-e0}. This fit is accurate to within 0.1--1\%
at $\gamma=10^{-1}-10^4$, and it also provides
the correct limits at $\gamma\to0$.

\begin{table}
\caption{Parameters of the analytical approximation 
\protect\eref{e0gamma} for the energies
of tightly-bound states $|0s0\rangle$
at $10^{-1}\leq\gamma\leq10^4$.}
\label{tab-e0}
\lineup
\begin{indented}
\item[]\begin{tabular}{@{}llllll}
\br
$s$ & $p_1$ & $p_2$ & $p_3$ & $p_4$ & $p_5$ \\
\mr
 0 & 15.55   & 0.378 & 2.727 & 0.3034 & 0.4380 \\
 1 & 0.5332  & 2.100 & 3.277 & 0.3092 & 0.3784 \\
 2 & 0.1707  & 4.150 & 3.838 & 0.2945 & 0.3472 \\
 3 & 0.07924 & 6.110 & 4.906 & 0.2748 & 0.3157 \\
 4 & 0.04696 & 7.640 & 5.787 & 0.2579 & 0.2977 \\
 5 & 0.03075 & 8.642 & 6.669 & 0.2431 & 0.2843 \\
 6 & 0.02142 & 9.286 & 7.421 & 0.2312 & 0.2750 \\
 7 & 0.01589 & 9.376 & 8.087 & 0.2209 & 0.2682 \\
\br
\end{tabular}
\end{indented}
\end{table}

For the hydrogen-like states, 
we use the asymptotic result (Haines and Roberts 1969)
\be 
\fl
   E^{(0)}_{ns\nu}={\mbox{1 Ryd}\over (n+\delta)^2},
  \quad \mbox{where}\quad\left\{
   \begin{array}{ll}
   n=(\nu+1)/2, ~~\delta\sim\gamma^{-1}\qquad & \mbox{(odd $\nu$)}\\
   n=\nu/2, ~~\delta\sim(\ln\gamma)^{-1}& \mbox{(even $\nu$)}.
   \end{array}
   \right.
\label{e0h}
\ee 
We have obtained the following fits to 
the quantum defect $\delta$: for odd $\nu$,
\be 
  \delta(\gamma)=
   (a_\nu+b_\nu\sqrt{\gamma}+0.077\gamma)^{-1},
\label{e0odd}
\ee 
where $a_\nu\approx1$ and $b_\nu\approx2$; and
for even $\nu$,
\be 
    \delta(\gamma)=
   \left[a_\nu+1.28\,\ln(1+b_\nu\gamma^{1/3})\right]^{-1},
\label{e0even}
\ee 
where $a_\nu\approx\case23$ and $b_\nu\approx\case23$.
Accurate values of $a_\nu$ and $b_\nu$
are given in \tref{tab-e0h}.
At $1\leq\gamma\leq10^4$, 
rms errors of \eref{e0odd} lie within $3\times10^{-4}$, 
and those of \eref{e0even} within $10^{-3}$.

\begin{table}
\caption{Parameters of the analytical approximations 
\protect\eref{e0h}--\protect\eref{e0even} for the energies
of hydrogen-like states $|00\nu\rangle$
at $1\leq\gamma\leq10^4$.}
\label{tab-e0h}
\lineup
\begin{indented}
\item[]\begin{tabular}{@{}lllllll}
\br
$\nu$          & 1     & 2     & 3     & 4     & 5     & 6     \\
\mr
$a_\nu$        & 0.785 & 0.578 & 0.901 & 0.631 & 0.970 & 0.660 \\
$b_\nu$        & 1.724 & 0.765 & 1.847 & 0.717 & 1.866 & 0.693 \\
\br
\end{tabular}
\end{indented}
\end{table}

The parameters 
$m_{\rm eff}$, $\alpha$ and $q_0$
in (\ref{eappr0})--(\ref{eappr3})
that determine $K$-dependences of the energies
can also be replaced by analytical functions of $\gamma$.
Let us start with the tightly-bound states ($\nu=0$).
For the effective mass, we have
\be 
   m_{\rm eff}(\gamma)=\mH\left[1+(\gamma/\gamma_0)^{c_0}\right],
\ee 
where the power index $c_0$ and the value $\gamma_0$
(roughly corresponding to the onset of strong coupling
between internal and centre-of-mass motions of the centred atom)
depend on the quantum number $s$ and are given by
$$
   c_0=0.937+0.038s^{1.58}
\mbox{~~~and~~~}
   \gamma_0 = 6150\,\left(1+0.0389s^{3/2}\right)
          \left[1+7.87s^{3/2}\right]^{-1}.
$$
For the critical pseudomomentum, 
we write
\be 
   q_0\equiv K_c/\sqrt{2\mH E^{(0)}} =
    c_1+\ln(1+\gamma/\gamma_1).
\label{q0}
\ee 
The parameters $c_1$ and $\gamma_1$ take on the values
$c_1=0.81,1.09,1.18,1.24$
and
$\gamma_1=(8.0,3.25,2.22,1.25)\times10^4$
for
$s=0,1,2,3$, respectively.
For $s\geq4$, we put $c_1=0.93+0.08s$ and $\gamma_1=6500$.
The remaining parameters can be
replaced by simple expressions,
$\alpha=0.053\,\ln(\gamma/150)$ and
$q_2=0.158\,[\ln((1+0.1s)\gamma/215)]^{2/5}$.

Now let us turn to the hydrogen-like states.
For odd states, we have, approximately,
$q_0=(\nu^{5/4}\gamma/170)^{0.9}$ and 
$\alpha=0.66+\nu/20$, whereas for even
hydrogen-like states
$q_0=\nu\,\sqrt{\gamma/1200}$ and
$\alpha=0.66$.

\begin{figure}[t]
\begin{center}
\leavevmode
\epsfysize=140mm
\epsfbox[30 180 480 670]{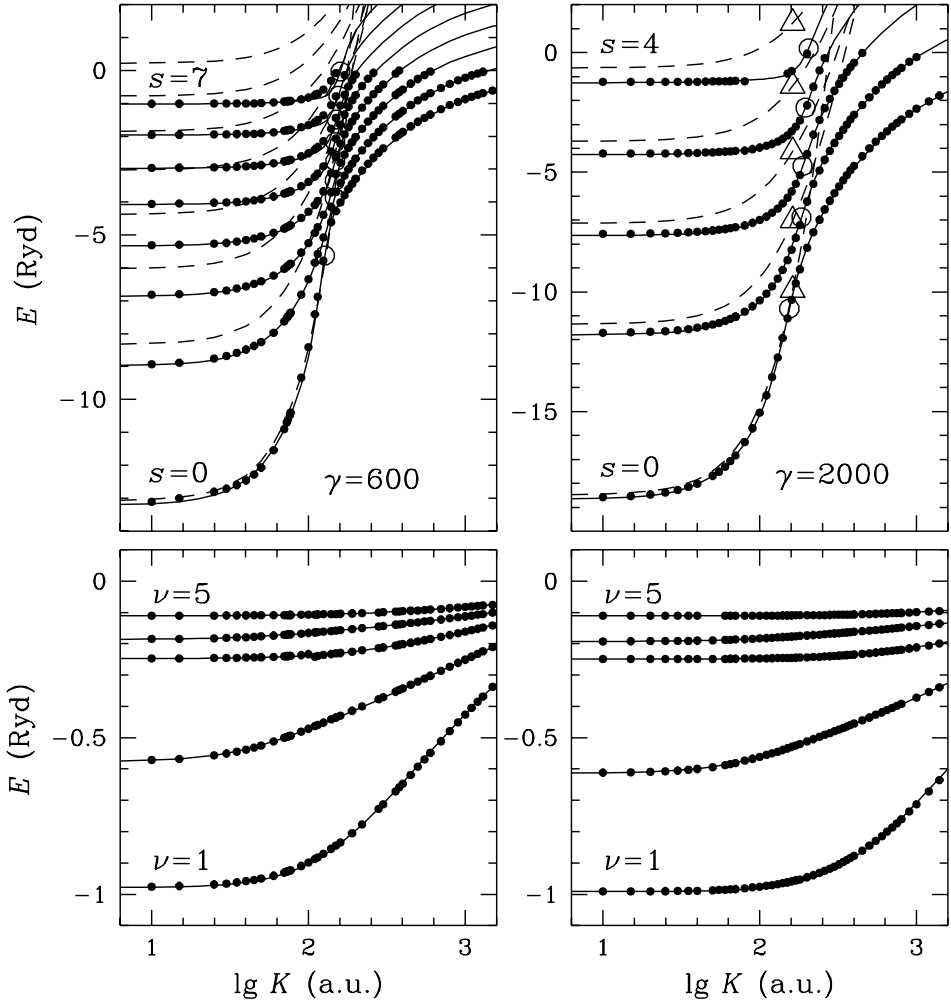}
\end{center}
\caption[]{
Energy spectrum of the hydrogen atom moving across
a strong magnetic field.
Upper panels: tightly-bound states ($\nu=0$);
lower panels: hydrogen-like states of the manifold $s=0$.
Numerical values (\protect\fullcirc) are compared with 
the analytical approximations 
of \protect\sref{sect-2dim} (\protect\full) 
and of Lai and Salpeter (1995) 
(\protect\broken, upper panels). Triangles ($\triangle$) mark 
the limit of validity of the
perturbation formalism according to equation (3.8) of 
Lai and Salpeter; large circles (\protect\opencirc) point
the present analytical approximation for the
critical pseudomomentum $K_c$. 
\label{fig1}
}
\end{figure}

These approximations
are not so accurate as those 
provided by tables \ref{tab1} and \ref{tab2},
but their advantage is that they may be used 
at arbitrary $\gamma$ in the range considered.
In figure~\ref{fig1} they are compared 
with our numerical results
and with the fitting formulae of Lai and Salpeter (1995). 
The figure demonstrates that the present approximations
are valid at any $K$ from 0 to infinity.
Noticeable discrepancies between our fitted and calculated data
occur only in narrow ranges of $K$ near anticrossings.

\subsection{The largest transverse velocity}
As an example of application of the above formulae
for binding energies, let us estimate
the maximum velocity of the atom, 
$v_{\rm max}=\max|\partial E/\partial K|$. It
can be alternatively interpreted as the maximum
transverse electric field ${\cal E}_{\rm mov}=v_{\rm max}B/c$
that could be applied to an atom at rest. 
A stronger electric field ${\cal E}>{\cal E}_{\rm mov}$
forces the atoms to move with velocities around 
the drift velocity of free charges in crossed fields,
$\vv{v}_{\rm drift}=c\,\vv{{\cal E}}\times\vv{B}/B^2$,
provided that ${\cal E}<{\cal E}_c=B=137.036\,\gamma$~au
(in conventional units, 
${\cal E}_c=2.998\times10^4\,[B/{\rm G}]{\rm~V~m}^{-1}$).
Still higher electric field, ${\cal E}>{\cal E}_c$,
cannot be counterbalanced by motion, hence
it causes Stark ionization. 

A numerical evaluation of $v_{\rm max}$
requires multiple calculation of derivatives 
$\partial E_\kappa/\partial K$ 
in the most complicated region $K\sim K_c$. 
Thus the analytical approximations can be most
helpful here. A reasonable approximation
is simply
$
   v_{\rm max,appr}=\partial |E|/\partial K
$
at $K=K_c$, where $E$ is given by 
\eref{eappr0} and $K_c$ by \eref{q0}.
\Tref{tab-vmax} presents 
$v_{\rm max,appr}$ obtained using this approximation
for the lowest tightly-bound states
along with 
$v_{\rm max,LS}$ given by equation (3.30) of Lai and Salpeter (1995)
and with 
$v_{\rm max}$ evaluated numerically.
The values listed in \tref{tab-vmax} 
(in atomic units of velocity, 1 au$=2188$
km s$^{-1}$) can be translated into those of the critical 
electric field, ${\cal E}_{\rm mov}=\gamma v_{\rm max}
\mbox{[au]}\cdot(5.14\times10^{11}$ V m$^{-1}$). 

\begin{table}
\caption{The largest transverse velocities (in au) 
in the lowest states $|0s0\rangle$: 
numerical values $v_{\rm max}$ compared with the present 
analytical approximation $v_{\rm max,appr}$ and with 
the approximation $v_{\rm max, LS}$ of Lai and Salpeter (1995).}
\label{tab-vmax}
\lineup
\begin{indented}
\item[]\begin{tabular}{@{}llllllll}
\br
$s$ & 0 & 0 & 0 & 0 & 1 & 2 & 3 \\
$\gamma$ 
& 300\0 & 1000 & 3000 & 10\,000
& 3000 & 3000 & 3000
\\
\mr
$v_{\rm max}$ 
& 0.0588 & 0.0479 & 0.0372 & 0.0253
& 0.0240 & 0.0198 & 0.0173 \\
$v_{\rm max,appr}$
& 0.0622 & 0.0467 & 0.0367 & 0.0279
& 0.0232 & 0.0187 & 0.0154\\
$v_{\rm max, LS}$ 
& 0.0812 & 0.0551 & 0.0374 & 0.0228
& 0.0174 & 0.0124 & 0.0100\\
\br
\end{tabular}
\end{indented}
\end{table}

\section{Geometrical characteristics and radiative transitions}
\label{sect-geom}
\subsection{Atomic sizes and dipole moments}
\label{sect-size}
Geometrical characteristics of an atom play an important role
in distribution of atoms over quantum states in a plasma
and in their contribution to the plasma absorption 
coefficients, since a ``size'' of an atom
may be used to approximately evaluate effects of 
destruction of atoms
caused by random charge distribution in the plasma
(e.g., Potekhin 1996).
The $K$-dependence of rms sizes is complicated
and can be non-monotonous near anticrossings.
However, the sizes usually
need not be known with high precision, that relieves the problem
of fitting. The accuracy level of the approximations
presented in this section is typically several percent. 

At $K=0$, the atom is axially symmetric, 
and its rms sizes along the Cartesian coordinates
can be approximated 
as $l_{x0}=l_{y0}\approx\ab\,\sqrt{(s+1)/\gamma}$ and 
\begin{eqnarray}
   l_{z0}\approx
      \left\{ 1/\sqrt{2}+1/\ln[\gamma/(1+s)]\right\}
      (\mbox{Ryd}/ E^{(0)})^{1/2}\,\ab\qquad& 
        (\nu=0)\label{lz0}\\
   l_{z0}\approx
       (1.6\,\mbox{Ryd}/E^{(0)})\,\ab 
         & (\nu\geq1).
\label{lz00}
\end{eqnarray} 

Let us consider an atom moving along $Oy$.
Both transverse sizes of the electron ``cloud''
remain approximately independent of $K$:
$l_x\approx l_y\approx l_{x0}$.
However, the atom acquires a constant dipole moment
$\vv{d}=e\langle\vv{r}_p-\vv{r}_e\rangle$
proportional to the mean proton-electron separation
$\bar{x}=|\langle\vv{r}_e-\vv{r}_p\rangle|$.
This separation is always smaller than $r_c$,
and it approaches $r_c$ at $K\gg K_c$.
With inaccuracy up to 10\%, at $\gamma\geq300$, 
\begin{eqnarray}
   \bar{x}/ r_c &\approx& 1-
   \left[1+0.015\gamma^2\,\sqrt{1+s}\,(E^{(0)})^{-4}\right]^{-1}
   \left[ 1+(K/K_c)^{1/\alpha}\right]^{-1}
\nonumber\\
  && - \left[1+0.004\gamma^2\,(E^{(2)}(K))^{-4}\right]^{-1}
   \left[ 1+(K_c/K)^{1/\alpha}\right]^{-1},
\end{eqnarray}
where $E^{(0)}$, $E^{(2)}$, $K_c$
and $\alpha$ are defined above.

The size of the electron cloud along the field 
is also affected by the motion. 
It can be described by the formulae
\begin{eqnarray}
\fl
   l_z = l_{z0} {\left[1+(1-\mH/m_{\rm eff}) (K/K_c)^2\right]^{1/2}
          \over 1+(K/K_c)^{1/\alpha}}
      + {l_{z2}\over 1+(K_c/K)^{1/\alpha}}\qquad & (\nu=0)
\\
   l_z = \left(l_{z0}^2+l_{z2}^2\right)^{1/2}
      & (\nu\geq1).
\end{eqnarray}
Here, $l_{z0}$ is the value at $K=0$ given by \eref{lz0}, \eref{lz00}, 
and
\[
l_{z2}=\sqrt{\nu+1/2}\,\left[r_\ast^3+(4.3+7\nu^2)r_\ast^2\right]^{1/4}
\]
has the correct asymptotics at $K\to\infty$ (Paper~I).

\begin{figure}[t]
\begin{center}
\leavevmode
\epsfysize=100mm
\epsfbox[50 220 500 660]{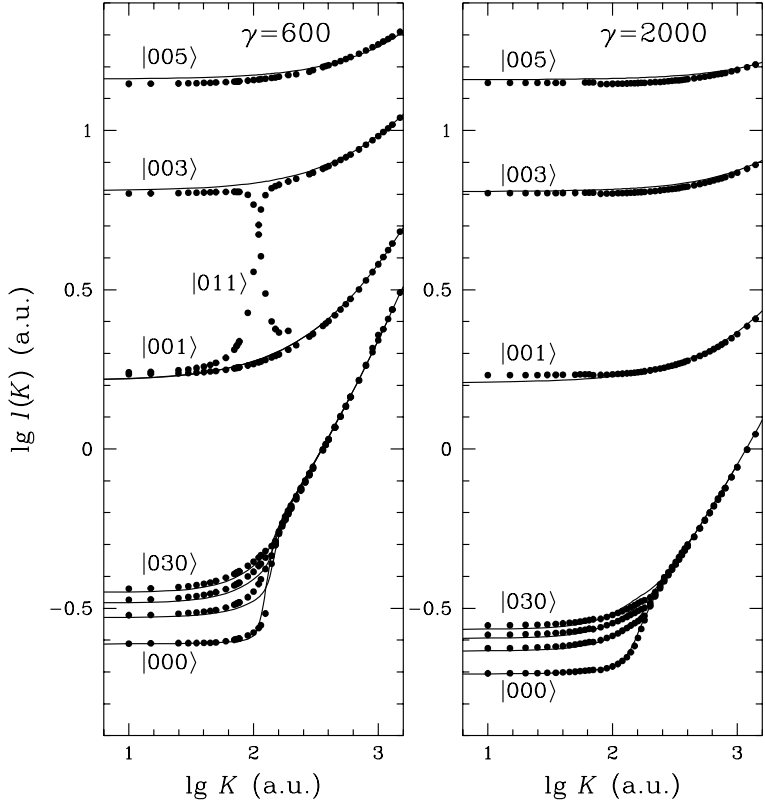}
\end{center}
\caption[]{
Comparison of calculated geometrical sizes of the atom 
(\protect\fullcirc) with the analytical approximations
of \protect\sref{sect-size} (\protect\full).
\label{fig2}
}
\end{figure}

In figure \ref{fig2} the average size of the atom, 
$l(K)=\left[\bar{x}^2+l_x^2+l_y^2+l_z^2\right]^{1/2}$,
expressed through the above formulae,
is compared with values calculated numerically.
On the left panel (at $\gamma=600$),
the strong deviations of the numerical values from the fit
for the states $|003\rangle$ and $|011\rangle$ at $K\sim10^2$~au
are caused by their anticrossing, which occurs shortly
before the level $(011)$ enters continuum.
At $\gamma=2000$ (right panel), this level
belongs to the continuum at arbitrarily small $K$,
so it does not (anti)cross truly bound levels.

\subsection{Oscillator strengths}
\label{sect-osc}
In this section we consider those
oscillator strengths $f$
that dominate photoabsorption 
of polarized radiation by ground-state hydrogen atoms at large $\gamma$.
The polarization is assumed circular 
(right, for which we will use superscript `+', or left, `$-$') 
or linear, longitudinal (`$\|$')
with respect to the static magnetic field.

At $K=0$, the left-polarized radiation cannot excite
the ground-state atom ($f^-=0$), while 
right and longitudinally polarized
radiation is absorbed mainly via transitions
to the states $|010\rangle$ and $|001\rangle$, respectively.
The corresponding oscillator strengths
have been computed and tabulated by Forster \etal (1984).
With an accuracy of 1--2\%, at $0\leq\gamma\leq10^4$,
they are reproduced by a single five-parameter formula
\be
   f^\alpha_{ns\nu}(0)=\left(1-{0.584\over
                             1+u_1\gamma^{u_2}}\right)
     {1+u_3\gamma \over 1+u_4\gamma^{u_5}}.
\label{oscfit0}
\ee
For $f^+_{010}(0)$, the parameteres $u_i$ take on the values
$u_1=12$, $u_2=1.43$, $u_3=9.8\times10^{-5}$, 
$u_4=1.585$ and $u_5=0.713$.
For $f^\|_{001}(0)$, we have 
$u_1=2.64$, $u_2=1.076$, $u_3=6\times10^{-6}$, 
$u_4=0.247$ and $u_5=0.381$.

For the moving atom, 
in the restricted range $300\leq\gamma\leq10^4$,
we put
\be
   f^+_{010}(K)=f^+_{010}(0)
     {1-a (K/K_c)^b\over 1+(K/K_c)^{1/\alpha'}}+
     {2(\mel/\mpr)\over 1+(K_c/K)^{1/\alpha'}},
\label{f010}
\ee
with $a=1.28-0.267\ln(1+\gamma/240)$,
$b=1+3/[1+\ln^2(1+\gamma/90)]$,
$\alpha'=0.012\,[1+\ln^2(1+\gamma/90)]$,
and
\be
   f^\|_{001}(K)=f^\|_{001}(0)\exp[-(a'K/K_c)^2]
   +{\exp[-(b'K/K_c)^{-\beta}]\over 1+0.5\sqrt{K_c/K}},
\label{f001}
\ee
with $a'=0.877\ln(13100/\gamma)$,
$b'=0.89-\gamma/17000$
and
$\beta=0.61\,(1+2410/\gamma)^{3/2}$.
The second parts of \eref{f010}, \eref{f001} 
ensure the correct large-$K$ limits ($2\mel/\mpr$
and 1 respectively, cf.\ Paper~I).

The radiative transitions forbidden for the atom at rest because of 
the conservation of the angular-momentum projection 
become allowed for the moving atom. In particular, 
the moving ground-state atom can absorb
left-polarized radiaton. Oscillator strengths
of such transitions are significant only at $K$
of the order of $K_c\sim10^2$ au.
Therefore we derive for them fitting formulae 
accurate to $\sim10$\% in this range of $K$
and do not attempt to fit the complicated behaviour
they show outside this range, where they are orders of magnitude
smaller (Paper~I). 

The transition 
to a state $|0s0\rangle$ presents
the dominating absorption channel for circular
polarization
in a spectral range
$E^{(0)}_{000}-E^{(0)}_{0,s-1,0}<\hbar\omega<
E^{(0)}_{000}-E^{(0)}_{0s0}$
(where $\hbar\omega$ is a photon energy).
For the right polarization, we put
\be
   f^+_{0s0}(K)={0.012\,(K/K_c)^{2s}(1-K/K_c)
    \over 1+11\ln(1+(\gamma/3300)^2)}
    \mbox{~~at $K<K_c$, $s\geq2$},
\ee
and zero at $K\geq K_c$;
for the left polarization,
\be
   f^-_{0s0}(K)={1.3\times10^{-4}(K/K_c)^{2(s+1)}
       \over 2^s[1+(K/K_c)^{5(s+1)}]}
    \qquad     (s\geq1).
\label{f0s0}
\ee
Although approximations \eref{f010}--\eref{f0s0} are rather crude,
particularly owing to the anticrossings,
their accuracy may be still sufficient for astrophysical applications,
as will be demonstrated in the next section.

\section{Spectral line shapes}
\label{sect-shape}
As an application of the above fitting formulae, 
let us consider bound-bound absorption spectrum
of hydrogen under the conditions typical for neutron star
atmospheres (Pavlov \etal 1995): density $\rho\gapr10^{-2}$ g cm$^{-3}$,
temperature $T\sim10^5-10^6$~K, and magnetic field strength
$B\sim10^{12}-10^{13}$~G. Such absorption spectra
have been studied in Paper~II.
Neglecting the Doppler and collisional broadening
but taking into account the most important magnetic broadening,
one obtains an average
partial cross section of an atom
with respect to absorption of polarized 
radiation with frequency $\omega$
via some specific transition $|\kappa\rangle\to|\kappa'\rangle$
as the sum
$
\sigma(\omega)=\sum_i \sigma_i(\omega)
$
over the roots $K_i(\omega)$ of the equation
$
E'(K)-E(K)=\hbar\omega,
$
where $E'(K)$ ($E(K)$) is the energy of the final (initial) 
state of the atom, and
\be
\sigma_i(\omega)=C_w^{-1}{4\pi^3e^2\over\mel c} 
K_i|{\rm d}K_i/{\rm d}\omega|
w'(K_i)\exp[-E(K_i)/\kb T] f(K_i).
\ee
Here, $f(K)$ is the oscillator strength for the given 
transition and polarization, 
$\kb$ is the Boltzmann constant,
\be
C_w=2\pi\int_0^\infty K w(K)\exp[-E(K)/\kb T] {\rm d}K
\ee
is a normalization constant, and $w'(K)$ ($w(K)$)
is the occupation probability of the final (initial)
atomic state
in a plasma with a number density of electrons $n_{\rm e}$.
These occupation probabilities can be estimated as
$
w\sim\exp[-(4\pi/3)n_{\rm e}(4l)^3],
$
where $l$ is the rms radius of the atom.
This expression is pertinent to calculation
of bound-bound absorption considered here.
For calculation of thermodynamic properties of the plasma,
however,
one should use, instead of these ``optical'' occupation
probabilities, ``thermodynamical'' ones, which are
generally larger (Potekhin 1996).

In order to take into account induced emission,
$\sigma_i(\omega)$ should be multiplied by
$(1-{\rm e}^{-\hbar\omega/\kb T})$.
The collisional broadening can be taken into account
by convolution of $\sigma_i(\omega)$ with the Lorentzian
profile characterized by the width $\Gamma(K_i(\omega))$
(Paper~II).
Since this type of broadening has only
marginal significance compared to the magnetic broadening
in the neutron star atmospheres, 
we employ a simple order-of-magnitude estimate:
\be
   \Gamma(K)\approx \Gamma_0 n_{\rm e}\ab^3 
   (\kb T/\mbox{Ryd})^{1/6}
   (1+2r_\ast^{5/6}),
\label{Gamma}
\ee
where $\Gamma_0\approx15$ au for 
transitions to the state $|001\rangle$ 
(that mainly determine absorption of radiation
polarized longitudinally)
and $\Gamma_0\approx(68/\gamma)$ au for transitions
to the state $|010\rangle$ (responsible for the main 
absorption peak of circular polarization).

\begin{figure}[t]
\begin{center}
\leavevmode
\epsfysize=90mm
\epsfbox[50 170 500 640]{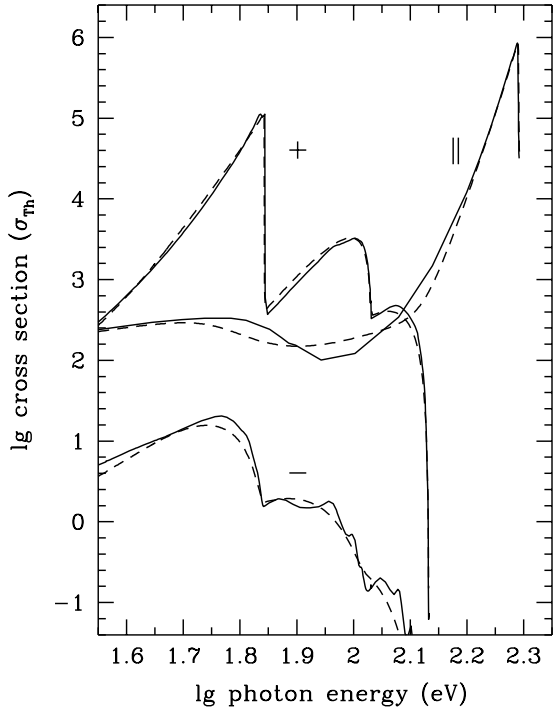}
\end{center}
\caption[]{
Comparison of numerically calculated (\protect\full)
and approximate (\protect\broken)
spectral shapes for bound-bound absorption
of right ($+$), left ($-$) and longitudinally
($\|$) polarized radiation by
hydrogen atoms in thermal equilibrium with
plasma under the conditions typical for 
atmospheres of magnetic neutron stars:
$\kb T=1$~Ryd ($T=1.58\times10^5$~K),
$\rho=0.01{\rm~g~cm}^{-3}$, and
$\gamma=10^3$ ($B=2.35\times10^{12}$~G).
\label{fig3}
}
\end{figure}

Some of the typical absorption profiles 
obtained numerically in Paper~II are represented
by full lines in figure \ref{fig3}.
The spectral range is shown that is relevant to interpretation
of spectral observations of neutron stars with
the X-ray telescope on-board {\it ROSAT\/} satellite (e.g.,
Pavlov \etal\ 1995).
Approximate profiles, obtained using the analytical approximations
of section \ref{sect-2dim} for $E(K)$,
the formulae of section \ref{sect-geom}
for $l(K)$ and $f(K)$,
and the estimate (\ref{Gamma}) for $\Gamma(K)$,
are shown
in figure \ref{fig3} by broken lines. 
One can see that they correctly reproduce
gross features of the spectral 
   shapes, and thus the proposed approximations are suitable for
   using them in theoretical models of neutron-star atmospheres
   compatible with contemporary observational data.
   The figure demonstrates also the importance
   of the decentred states: for example, 
   the absorption of longitudinally
   polarized radiation at $\hbar\omega<100$~eV is produced solely 
   by the states of motion with $K>K_c$.

In reality, observed spectra are influenced not only by 
the bound-bound photoabsorption shown in figure \ref{fig3}
but also by bound-free transitions.
For moving atoms, bound-free
cross sections were calculated
by Potekhin and Pavlov (1997).
Under conditions typical of atmospheres of cooling neutron stars,
maximum bound-free photoabsorption
turns out to be of the same order of magnitude
as the bound-bound one, but it is shifted to 
higher photon energies. In the spectral range
shown in figure \ref{fig3}, the bound-free transitions
are generally less important.

\section{Conclusions}

We have obtained analytical approximations
of binding energies, geometrical sizes
and main oscillator strengths of radiative transitions
of the hydrogen atom moving across a strong magnetic field.
   These approximations can be also applied
   to the hydrogen atom in crossed electric and magnetic fields,
   since the latter problem reduces to the former one
   with the effective pseudomomentum 
   $K=\mH c \,{\cal E}/{\cal E}_c$, or equivalently
   $K{\rm~[au]} = 8.4 \,({\cal E}/{\rm V~m}^{-1})\,(B/{\rm G})^{-1}$.

Binding energies are the most important
quantities in many applications,
and for that reason we have presented not only
fitting formulae
analytically depending on $\gamma$ and $K$ (section \ref{sect-2dim}),
but also considerably more accurate $K$-dependences 
at 6 selected values  of $\gamma$ (section \ref{sect-1dim}).
Atomic sizes (section \ref{sect-size}) play important role
in distribution of atoms over quantum states in a plasma
and in their contribution to the plasma absorption 
coefficients. For example, a size of an atom
may be used to evaluate effects of ``unbinding''
of electrons caused by random charge distribution in the plasma.
For non-magnetized hydrogen plasmas, an approximate treatment of
these effects was revised recently (Potekhin 1996);
for the strong magnetic fields analogous work is under way.
The approximations of oscillator strengths (section \ref{sect-osc}),
along with those of the energies and sizes,
facilitate calculations of absorption spectra
of strongly magnetized, partially ionized hydrogen plasmas.
Eventually, the analytical estimates of $\gamma$- and 
$K$-dependences of the 
binding energies, atomic sizes and transition rates will
help to generalize previously developed models of
fully ionized atmospheres of magnetic neutron 
stars (Shibanov \etal 1992)
to the more realistic case of partially ionized atmospheres.

\ack
This work was inspired by 
stimulating discussions with participants 
of the 172 WE-Heraeus-Seminar 
on Atoms and Molecules in Strong External Fields 
(Bad Honnef, 7--11 April 1997),
organized by Dr.\ Peter Schmelcher and Dr.\ Wolfgang Schweizer.
I wish to express my gratitude to Prof.\ Joachim Tr\"umper and
Dr.\ Vyacheslav Zavlin 
for hospitality during my visit
at the Max-Planck-Institut f\"ur Extraterrestrische Physik
in Garching, where a large part of this work has been performed
using the MPE computing facilities.
The visit was made possible due to
the DFG--RFBR grant No.\,96-02-00177G.
Partial support from the RFBR grant No.\,96--02--16870a
and INTAS grant No.\,94-3834 is also acknowledged.
I am grateful to Prof.\ Joseph Ventura
for useful discussions and hospitality
at the University of Crete, where the final
part of this work was completed.
\newpage
\section*{References}
\begin{harvard}

\item[]
Avron J E, Herbst I W and Simon B 1978 {\it Annals of Phys. (N.\,Y.)} 
{\bf 114} 431--51
\item[]
Baye D and Vincke M 1990 \PR A {\bf 42} 391--6
\item[]
Baye D, Clerbaux N and Vincke M 1992 \PL A {\bf 166} 135--9
\item[]
Braun P A and Solov'ev E A 1984 \JPB {\bf 17} L211--6
\item[]
Burkova L A, Dzyaloshinskii I E, Drukarev G P and Monozon B S 1976 
{\it Sov. Phys.--JETP} {\bf 44} 276--8 
\item[]
Chen Z and Goldman S P 1992 \PR A {\bf 45} 1722--31
\item[]
Dippel O, Schmelcher P and Cederbaum L S 1994 \PR A {\bf 49} 4415--29
\item[]
Dzyaloshinskii I E 1992 \PL {\bf 165A} 69--71
\item[]
Elliott R J and Loudon R {\it J.\ Phys.\ Chem.\ Solids} {\bf 15} 196--207
\item[]
Fassbinder P and Schweizer W 1996
{\it Astron.\ Astrophys.} {\bf 314} 700--6
\item[]
Forster H, Strupat W, R\"{o}sner W, Wunner G, Ruder H and Herold H 
1984 \JPB {\bf 17} 1301--19
\item[]
Gorkov L P and Dzyaloshinskii I E 1968 {\it Sov. Phys.--JETP} {\bf 26} 
449--58
\item[]
Haines L K and Roberts D H 1969 {\it Am.\ J.\ Phys.} {\bf 37} 1145--54
\item[]
Ivanov M V 1988 \jpb {\bf 21} 447--62
\item{}
Ipatova I P, Maslov A Y and Subashiev A V 1984 
{\it Sov.\,Phys.--JETP} {\bf 60} 1037--9
\item[]
Johnson B R, Hirschfelder J O and Yang K-H 1983 \RMP 
{\bf 55} 109--53
\item[]
Lai D and Salpeter E E 1995 \PR A {\bf 52} 2611--23
\item[]
Klaassen T O, Dunn J L and Bates C A 1997
{\it Proc.\ 172 WE-Heraeus-Seminar
``Atoms and Molecules in Strong External Fields'' 
(Bad Honnef, 7--11 April 1997)},
ed P Schmelcher and W Schweizer (New York: Plenum) in press
\item[]
Kravchenko Yu P, Liberman M A and Johansson B 
1996 \PR A {\bf 54} 287--305
\item[]
Melezhik V S 1993 \PR A {\bf 48} 4528--38
\item[]
Pavlov G G and M\'{e}sz\'{a}ros P 1993 {\it Astrophys.\ J.} 
{\bf 416} 752--61
\item[]
Pavlov G G and Potekhin A Y 1995 
{\it Astrophys.\ J.} {\bf 450} 883--95 (Paper~II)
\item[]
Pavlov G G, Shibanov Yu A, Zavlin V E and Meyer R D 1995
{\it Proc.\ NATO ASI} C {\bf 450} 
{\it The Lives of the Neutron Stars},
ed M A Alpar, \"U Kizilo\u{g}lu and J van Paradijs
(Dordrecht: Kluwer) p~71--90
\item[]
Potekhin A Y 1994 \jpb {\bf 27} 1073--90 (Paper~I)
\item[]
Potekhin A Y 1996 
{\it Phys. Plasmas} {\bf 3} 4156--65
\item[]
Potekhin A Y and Pavlov G G 1993 {\it Astrophys.\ J.} {\bf 407} 330--41
\item[]
Potekhin A Y and Pavlov G G 1997 {\it Astrophys.\ J.} {\bf 483} 414--25
\item[]
Ruder H, Wunner G, Herold H and Geyer F 1994
{\it Atoms in Strong Magnetic Fields} 
(Berlin: Springer-Verlag)
\item[]
R\"osner W, Wunner F, Herold H and Ruder H 1984 \JPB {\bf 17} 29--52
\item[]
Shibanov Yu A, Zavlin V E, Pavlov G G and Ventura J 1992
{\it Astron.\ Astrophys.} {\bf 266} 313--20
\item[]
Schmelcher P 1993 \PR B {\bf 48} 14642--5
\item[]
Taylor J H, Manchester R N and Lyne A G 1993 
{\it Astrophys.\ J.\ Suppl.\ Ser.} {\bf 88} 529--68
\item[]
Vincke M and Baye D 1988 
{\it J.\ Phys.\ B: At.\ Mol.\ Opt.\ Phys.} {\bf 21} 2407--24 
\item[]
Vincke M, Le Dourneuf M and Baye D 1992 {\it J. Phys. B: At. Mol. Opt. 
Phys.} {\bf 25} 2787--2807
\item[]
Wintgen D and Friedrich H 1986 \JPB {\bf 19} 991--1011
\item[]
Wunner G and Ruder H 1987 {\it Phys. Scr.} {\bf 36} 291--9
\item[]
Xi J, Wu L, He X and Li B 1992 {\it Phys. Rev} A {\bf 46} 5806--11

\end{harvard}
\end{document}